\documentstyle[12pt,aasms4]{article}
\def\msunend{$M_{\sun}$}
\def\msun{$M_{\sun}$\thinspace}
\def\dash{------}
\def\yrm1{yr$^{-1}$}
\begin{document}
\title{Hydrogen-Accreting Carbon-Oxygen White Dwarfs of Low Mass: 
Thermal and Chemical Behavior of Burning Shells}

\author{Luciano Piersanti  \altaffilmark{1}}
\affil{Dipartimento di Fisica dell'Universit\`a degli Studi di Napoli
``Federico II'', Mostra d'Oltremare, pad. 20, 80125, Napoli, Italy; 
lpiersanti@astrte.te.astro.it}

\author{Santi Cassisi}
\affil{Osservatorio Astronomico di Teramo, Via M.Maggini 47, 
64100 Teramo, Italy; cassisi@astrte.te.astro.it} 

\author{Icko Iben Jr.}
\affil{Astronomy and Physics Departments, University of Illinois, 
1002 W. Green Street, Urbana, IL 61801;icko@astro.uiuc.edu}

\and

\author{Amedeo Tornamb\'e \altaffilmark{2}}
\affil{Osservatorio Astronomico di Teramo, Via M.Maggini 47, 
64100 Teramo, Italy; tornambe@astrte.te.astro.it} 

\noindent
\altaffilmark{1}{Osservatorio Astronomico di Teramo, Via M.Maggini 
47, 64100 Teramo, Italy} 
\altaffilmark{2}{Dipartimento di Fisica, Universit\`a de L'Aquila,
Via Vetoio, 67100 L'Aquila, Italy}

\begin{abstract}
\noindent
Numerical experiments have been performed to investigate the thermal 
behavior of a cooled down white dwarf of initial mass $M_{\rm WD} = 
0.516 M_{\sun}$ which accretes hydrogen-rich matter with $Z = 0.02$ 
at the rate $\dot{M}=10^{-8}$ \msun \yrm1, 
typical for a recurrent hydrogen shell flash regime. The evolution 
of the main physical quantities of a model during a pulse cycle is 
examined in detail. From selected models in the mass range 
$M_{\rm WD} = 0.52\div 0.68$ \msunend, we derive the borders in the 
$M_{\rm WD}$ - $\dot{M}$ plane of the steady state accretion regime when 
hydrogen is burned at a constant rate as rapidly as it is accreted. 
The physical properties during a hydrogen shell flash in  white dwarfs
accreting hydrogen-rich matter with 
metallicities $Z = 0.001$ and $Z = 0.0001$ are also studied. For a 
fixed accretion rate, a decrease in the metallicity of the accreted 
matter leads to an increase in the thickness of the hydrogen-rich
layer at outburst and a decrease in the hydrogen-burning shell efficiency.
In the $M_{\rm WD}$-$\dot{M}$ plane, the borders of the steady state
accretion band are critically dependent on the metallicity of the
accreted matter: on decreasing the metallicity, the band is shifted to
lower accretion rates and its width in $\dot{M}$ is reduced. 
\end{abstract}

{\em Subject headings}: stars: novae, cataclysmic 
variables - stars: accretion - supernovae: general - white dwarfs  

\section{Introduction}

\noindent
The knowledge of the physical consequences of the accretion of 
hydrogen-rich matter 
onto a white dwarf has played an important role in understanding 
the main properties of several types of eruptive stars such as 
slow and fast novae and symbiotic stars (e.g., Starrfield 1971, 
Starrfield, Truran, \& Sparks 1978, Sparks, Starrfield, \& Truran 
1978). In addition, the scenario in which a red giant star transfers 
mass to its carbon-oxygen (CO) white dwarf companion (Whelan \& 
Iben 1973) is regarded by some as one of the most plausible 
precursor candidates for Type Ia supernovae (e.g., Hachisu, Kato, \& 
Nomoto 1996). Fairly extensive surveys of the dependence of behavior 
on white dwarf mass and accretion rate have been conducted (see, 
e.g., Iben \& Tutukov 1996; Cassisi, Iben, \& Tornamb\'e 1998
[hereinafter CIT] and references therein). For a low mass CO white
dwarf of typical mass in the range $0.5\div 0.8$ \msunend, the
consequences of the accretion of hydrogen-rich matter of solar
metallicity can be summarized as follows:
\begin{itemize}
\item for sufficiently large mass-accretion rates (say, $10^{-7}$ 
\msun \yrm1 or larger, depending on the white dwarf mass), the 
accreted layer adopts an expanded configuration similar to that of 
the envelope of a red giant star;
\item for intermediate mass-accretion rates (say, in the range $4\div 10
\times 10^{-8}$ \msun \yrm1), the accretor burns hydrogen in a 
steady state regime at the same rate as it accretes hydrogen;
\item for small mass-accretion rates (say, in the range $1\div 4 \times 
10^{-8}$ \msun \yrm1), the accretor experiences recurrent mild 
flashes;
\item for even smaller mass-accretion rates (say, smaller than 
$10^{-9}$ \msun \yrm1), the accretor experiences very strong 
nova-like hydrogen shell flashes.
\end{itemize}

Although considerable attention has been paid to models that 
experience recurrent mild hydrogen-burning flashes (e.g., 
Paczy\'nski \& $\dot{Z}$ytkow 1978, Iben 1982; Jos\'e, Hernanz, \& 
Isern 1993, CIT), a deep insight into the evolution 
of the main physical properties in the accreting models 
over an outburst-cooling cycle is still missing. 
Analytical studies of hydrogen-burning shells by Sugimoto \& 
Fujimoto (1978) and by Fujimoto (1982a,b) have established the 
general properties of the hydrogen-burning shell as a function of the 
fundamental parameters of the accreting star, but have not provided 
profiles of structural and chemical variables in the shell itself. 
Finally, the extant numerical experiments do not explore 
systematically how the outcome of the accretion process depends on 
the abundances of heavy elements in the accreted matter.

To investigate in detail the evolution of the main physical 
characteristics of the hydrogen-burning shell during a flash episode,
we have adopted as an initial model a white dwarf of mass 0.516 \msun
which has accreted matter at the rate $\dot{M} = 10^{-8}$ \msun \yrm1
for $7.6\times 10^{5}$ yr, at which point the interior has been cooled
to a temperature of $8.6\times 10^{6}$ K and the density at the
center is $2.56 \times 10^{6}$ g cm$^{-3}$. For the composition of
accreted matter, we have adopted a helium abundance by mass of $Y=0.28$
and three different metallicities ($Z=0.02$, 0.001, and 0.0001).

In \S 2 we discuss the input physics and the assumptions. In \S 3, the
thermal properties of the hydrogen-burning shell are presented and
discussed in detail for the Z=0.02 case. In \S 4, analytical relations
in the ($M_{\rm WD}$-$\dot{M}$) plane are obtained for the case Z=0.02
and, in \S 5, the dependence on metallicity of the evolutionary behavior
of the accreting models is analyzed. Conclusions and a brief discussion
follow in \S 6.


\section{Input Physics and Assumptions}

\noindent
The initial cold white dwarf model and the accretion experiments 
have been computed with an updated version of the FRANEC code 
(Chieffi \& Straniero 1989). A detailed discussion of the main 
differences with respect to the original version of the code is
given in CIT. The initial model is the same as that used in CIT 
and a description of the pre-accretion properties and the 
first pulse episode can be found in CIT. The main properties
of this model are listed in Table 1.

The accretion process is computed on the 
assumption that the accreted matter and the white dwarf surface have 
the same specific entropy; that is, it is assumed that all of the 
energy liberated by matter as it falls onto the surface of the white 
dwarf is radiated away.

The input physics differs from that used in CIT only in the low
temperature opacities: for $Z=0.02$ and $Z=0.0001$, we adopt the
opacity tables provided by Cox \& Stewart (1970a,b) and by
Cox \& Tabor (1976); for $Z=0.001$, the opacity values provided by 
Alexander \& Fergusson (1994) have been used. The initial
distribution of heavy elements adopted for models of
metallicities Z=0.02 and Z=0.0001 is the solar one as given by
Ross \& Aller (1976); for Z=0.001, the initial distribution is taken
from Grevesse (1991).

\placetable{table1}


\section{A Typical Hydrogen-Flash-Driven Pulse Cycle}

\noindent
In this section, we analyze the thermal and nuclear evolution of the 
hydrogen-burning shell of a model which has experienced enough (about 60) 
pulses that pulse properties have reached a local asymptotic limit. 
The model accretes hydrogen-rich matter of solar metallicity 
(Z=0.02) and its total mass at this point has grown to $M_{\rm 
WD}\sim 0.5236 M_{\odot}$. We assume that no mass is lost by the 
accreting star, despite the fact that a relatively nearby companion 
is required to supply the hydrogen-rich matter to the white dwarf 
and that, during evolution at high luminosity and low surface 
temperature, the surface of the star in outburst, in the real 
analogue, extends in most instances beyond the Roche lobe of the 
accretor and even beyond the donor star.

The most relevant structural properties of this model are reported in
Table 1. The well known evolution in the HR diagram during one complete pulse 
cycle is reported in Figure 1. Several points of special interest are
noted along the track: the positions where flash-driven convection
begins (IC) and where it disappears (EC); the positions where maxima
and minima of $\Phi_{\rm H} = L_{\rm H}/L_{\rm s}$ and $\Phi_{\rm He}
= L_{\rm He}/L_{\rm s}$ occur. Here, $L_{\rm H}$, $L_{\rm He}$, and
$L_{\rm s}$ are, respectively, the hydrogen-burning luminosity, the
helium-burning luminosity, and the surface luminosity. Several interior
characteristics during two passages of the cycle are shown in Figure 2
as a function of the total mass of the accretor.

\placefigure{fig1}

We begin our description at the upper right hand portion of the 
evolutionary track as the model evolves from red to blue along the 
high luminosity plateau and $\Phi_{\rm H} \sim 0.994$. In this phase, the 
release of gravothermal energy $\Phi_{\rm gr} = L_{\rm gr}/L_{\rm s}$ is equal 
to $\sim 0.006$, whereas the Helium burning contributes negligibly to the 
surface luminosity ($\Phi_{\rm He} << 1$).
Figure 2a discloses that, along the plateau 
portion of the evolutionary track, the location in mass of the 
hydrogen-burning shell (which we define as the point where the 
maximum rate of energy production via hydrogen-burning is located) 
moves outwards much more quickly than the total mass grows due to 
the accretion of fresh matter. In the next section, it will be shown how,
during the plateau phase, the relationship between the surface luminosity 
and the mass of the hydrogen-exhausted core depends on the metallicity
of the accreted matter.

\placefigure{fig2}

Once the mass of the layer between the hydrogen-burning shell and 
the surface decreases below a critical value, energy production by 
hydrogen burning declines rapidly (the reasons for this are 
discussed by Iben [1982] and the observational consequences are 
discussed by Iben \& Tutukov [1984]). The critical point in the HR 
diagram is the point of maximum effective temperature labeled BP 
(for ``blue point'') in Figure 1.

After reaching the blue point, the hydrogen-burning efficiency 
plumets and gravothermal energy takes over as the main source of 
energy (Figure 2d). In the absence of mass 
accretion, further evolution would be similar to that of a single 
star after leaving the AGB to become the central star of a planetary 
nebula; i.e., apart from the onset of crystallization, evolution 
would not have presented any additional curiosities. However, 
continuous mass accretion leads to a quite different behavior for the
outer layers of the white dwarf relative to that of a non-accreting,
cooling white dwarf. In particular, the combined action of the 
accretion process (growth in mass of the hydrogen-rich envelope) and 
the contraction of the layers above the hydrogen-burning shell (see 
Figs. 2e and 2c) relatively quickly brakes the rate of decline of 
the temperature of the hydrogen-burning shell (Fig. 2b), and leads 
to a slowly, but continuously, growing hydrogen-burning luminosity 
(Fig. 2d). 
In addition, it is worth noticing that the accretion process induces 
an increase of the evolutionary lifetime in the blue side of the HR loop 
from the bluest point to the faintest one. In particular along this portion 
of the cycle the accreting model evolves in a 50\% longer time-scale.

The energy delivered by hydrogen burning adds to the rate 
at which the local temperature in the shell increases until 
eventually a new hydrogen shell flash occurs. The increase in 
pressure related to the increase in temperature in the 
hydrogen-burning shell leads to a rapid expansion of the layers 
above the shell (see the surface radius increase in Fig. 2f and the 
decrease in density at the center of the shell in Fig. 2c). In the 
HR diagram, the model evolves upward as energy diffuses 
from the burning shell to the surface, and the increase in the 
surface radius eventually causes the model to evolve to the blue 
until the plateau portion of the track is reached once again.

Thus, the accretion process modifies the thermal content of the
hydrogen-rich layer, producing the physical conditions suitable for 
a new hydrogen shell flash. An examination of the evolution of the 
temperature profile in the outer layers of the model and of the 
profile of the rate of nuclear energy generation in the hydrogen-rich 
layer shows how this takes place. In Figure 1, circles, triangles, and
squares indicate the locations of all models for which profiles are given 
explicitly in Figures 3 and 4. All models marked by a given 
symbol in Figure 1 are represented by profiles plotted in a specific 
panel in Figures 3 and 4. Models marked by 
solid disks in Figure 1 are represented by the temperature profiles 
in Figure 3a and the energy-generation profiles in Figure 4a. Models 
marked by open triangles in Figure 1 are represented by profiles 
in Figures 3b and 4b. This continues in a counter clockwise fashion with, 
eventually, the models designated by open squares in Figure 1 being 
represented by the profiles in Figures 3f and 4f.

\placefigure{fig3}

\placefigure{fig4}

The temperature profile in Figure 3a which has the narrowest 
``peak'' (and acts as a lower envelope of the ensemble of profiles) 
describes the model designated by the solid disk at smallest surface 
temperature in Figure 1. Thus, as the model evolves in the 
HR diagram along the plateau branch from red to blue, thermal energy 
in the model interior flows inward in the form of a ``thermal 
wave,'' heating up hydrogen-free matter. At the same time, thermal 
energy diffuses outward from the hydrogen-burning shell, which 
itself propagates outward in mass toward the model surface; the 
temperature profile between the shell and the surface steepens.

During the next portion of the evolution (the open triangles in Fig. 
1 and the profiles in Fig. 3b), thermal energy stored over a fairly 
large fraction of the outer layers of the model (clearly beyond 
$M_{\rm r} \sim 0.523$ \msun in Fig. 3b) leaks outward and the drop 
in temperatures in the hydrogen-burning shell is reflected in a 
decline in the hydrogen-burning luminosity. Matter interior to 
$M_{\rm r} \sim 0.522$ \msun is still being heated from above. This 
behavior continues for the next designated set of models (solid 
boxes in Fig. 1 and profiles in Fig. 3c), with the ``watershed'' for 
thermal energy flow (the place from which energy flows both inward 
and outward) moving inward. For the following set of models (open 
circles in Fig. 1 and profiles in Fig. 3d), the energy-flow 
watershed has moved into interior regions not shown and, until the 
last few models, when hydrogen burning is beginning to make itself 
evident again (the bumps near the surface in the temperature 
profiles in Fig. 3d), cooling prevails over the entire region above 
$M_{\rm r} \sim 0.521$ \msunend.

As the new hydrogen-burning shell flash gets underway (the last 
three open circles and the closed triangles in Fig. 1), a large 
fraction of the nuclear energy which is released is stored locally, 
since the time scale on which thermal energy can be transferred is 
greater than the time scale on which nuclear energy is released. 
This is true even though matter in the burning region is not 
degenerate. A convective shell is formed early on in the development 
of the flash (the point labeled IC in Fig. 1). The mass of the 
convective shell grows until the shell extends from the base of the 
burning shell up to photospheric layers. Up to this moment, the 
evolution of the model has occurred on the nuclear burning time scale.
Thereafter, the convective shell (now, really, the convective envelope)
cannot accommodate a further increase in its thermal energy content,
but must expand to giant dimensions, using up local thermal energy to
do the work of expansion against gravity. The readjustment to a new,
expanded configuration occurs on the thermal time scale of the envelope.
When the energy surplus has been dissipated, (excursion to higher 
luminosity and lower effective temperature) and the envelope has 
readjusted to an expanded configuration, the thermal wave begins to 
propagate inward again (see Fig. 3f). The model is now back to where 
it began (the profile in Fig 3a with the narrowest peak) .

The profiles of the nuclear energy-generation rate $\epsilon_{\rm 
n}$ in Figure 4 are also instructive. As it is evident from Figure 4a, 
along the high luminosity plateau branch, the width in mass of the 
nuclear energy-generating region increases slightly and the 
hydrogen-burning luminosity drops as the burning shell works its way 
toward the surface. During most of the plateau phase, the full CNO 
cycles are active. By the time the model has attained its maximum 
effective temperature (Fig. 4b), hydrogen burning rapidly declines in 
importance as temperatures in the shell (Fig. 3b) decrease. The 
decline continues along the next portion of the cooling phase (Fig. 
4c) as temperatures in the shell continue to decrease (Fig. 3c). An 
interesting aspect of nuclear burning during the long cooling phase 
(Figs. 4b, 4c, and 4d) is that it takes place in two distinct 
regions: a left hand region where the CNO cycles operate and a right 
hand spike where $^{12}$C in freshly accreted fuel is burned into 
$^{14}$N. Along this phase the pp contribution to energy production 
plays a not negligible role (see the broad secondary peak among the 
spikes in Fig. 4d).

Ultimately, thanks to the increase in density at the base of the 
hydrogen-rich layers, heating replaces cooling in the 
hydrogen-burning layers (Figs. 3d and 4d) and both the CNO 
cycles and pp-driven $\epsilon_{\rm n}$ profiles begin to 
increase in height and in breadth. When convection appears, the 
second $\epsilon_{\rm n}$ peak is ``swallowed'' by the first (Fig 
4d). After the hydrogen-burning luminosity has attained its maximum 
value and convection begins to recede (Fig. 4e), the mass-width of 
the region undergoing the strongest hydrogen burning decreases, 
attaining its minimum width (Fig. 4f) at the same time the model reaches 
in the HR diagram the minimum effective temperature. 


\section{Steady-State Regime for Small White Dwarf Masses}

\noindent
The evolution described in the previous section illustrates the well 
known result that, for accretion rates smaller than a critical value
which depends on white dwarf mass, accreting white dwarfs can be 
viewed as evolving alternatively in two stable states (a low state and a 
high state, in the nomenclature of Fujimoto [1982b]), separated by 
short lived transitional phases. The low state is the cooling phase, 
when the gravothermal energy source supplies essentially the entire 
surface luminosity, with hydrogen burning being almost extinguished. 
The high (or ``excited'') state is the high luminosity plateau phase 
during which the hydrogen-burning shell is the main energy source, 
and the contribution of gravothermal energy is minor.

The hydrogen shell flash acts as the excitation mechanism which 
induces the transition from the low to the high state. The trigger 
for the transition is that, when the mass of the accreted layer 
exceeds a critical value which depends on the accretion rate, the 
rate of local heating by the hydrogen-burning shell exceeds the rate 
at which heat can diffuse out, initiating a thermonuclear runaway 

The ``strength'' of a flash (the maximum hydrogen-burning 
luminosity) and the maximum extension to the red of the evolutionary 
track during the excited 
state depend on the accretion rate, in the sense of being greater, 
the smaller the accretion rate, but the final plateau luminosity 
during the evolution from red to blue depends only on the mass of 
the white dwarf. Since the amount of mass accreted between flashes 
is a quite small fraction of the total mass, the white dwarf mass 
may be thought of as constant over a large number of cycles; to a
very good approximation, the structure of the envelope during the 
plateau phase does not ``remember'' the accretion rate preceding the 
flash which produced it, being sensitive only to the mass of the 
underlying white dwarf.

The transition from the excited to the low state sets in at the blue 
point along the evolutionary track in the HR diagram (point BP in 
Fig. 1). The sudden drop in $\Phi_{\rm H}$ that is shown in 
Figure 2d is initiated at the blue point. The reason for this second 
transition can be understood in terms of the dependence on accretion 
rate of the mass of the hydrogen-rich layer $\Delta M_{\rm H}$ in 
static models which are forced to burn hydrogen at the same rate as 
they accrete it (Iben 1982). The blue point in the HR diagram of the 
locus formed by a sequence of static models of fixed mass but 
different accretion rate is a bifurcation point (see Fig. 2 in Iben 
1982), such that models of successively higher luminosity and lower 
surface temperature than at the blue point have larger $\Delta 
M_{\rm H}$, whereas models with successively lower luminosity and 
surface temperature also have larger $\Delta M_{\rm H}$. If one 
imagines turning off the accretion rate in any particular model on 
the upper branch and letting this model evolve statically, $\Delta 
M_{\rm H}$ in that model would decrease because of nuclear burning 
and the model would evolve stably to the blue along the sequence, 
arriving at positions occupied by static steady state models of the 
same (successively smaller) $\Delta M_{\rm H}$. 

If, however, mass accretion were switched off in a model along the lower 
branch and this model were allowed to evolve statically, the model 
would evolve upward until it reached the bifurcation point, at which 
position it would be faced with a quandary. It could not evolve 
statically in either direction from the bifurcation point.

The implication of these thought experiments is that a real star 
evolving from red to blue along the plateau branch can do so in a 
roughly static fashion, with $\Phi_{\rm H} \sim 1$ until, on 
reaching the bifurcation point, hydrogen burning can no longer 
control the course of evolution. The static approximation is no 
longer valid, $\Phi_{\rm H}$ plumets until hydrogen burning is no 
longer of significance and gravothermal energy has taken over as the 
prime source of surface luminosity.

These considerations allow us to make use of the properties of 
accreting models in the quasistatic approximation to estimate, for 
small white dwarf masses, the lower boundary of the region in the 
$M_{\rm WD}$-$\dot{M}$ plane where steady state accretion can occur. 
In Figure 5, the rate $\epsilon_{\rm H}$ of energy 
generation at the center of the hydrogen-burning 
shell (panel a) and the rate $\dot{M}_{\rm H}$ at which the 
hydrogen-burning shell processes matter (panel b) in a model of mass 
0.5236 \msun accreting $Z = 0.02$ matter at the rate $10^{-8}$ \msun 
\yrm1 are shown. The position at the maximum surface temperature occurs along 
the curves in both panels corresponds to the bluest point in the HR 
diagram of Figure 1, and, there, $\log{\dot{M}_{\rm H}} = -7.554$ 

\placefigure{fig5}

In the $M_{\rm WD}$-$\dot{M}_{\rm H-sh}$ plane of Figure 6 are shown 
curves formed by four additional sets of models for larger white 
dwarf masses but for the same accretion rate of $10^{-8}$ \msun \yrm1.
The values of $\dot{M}_{\rm H}$ at the blue points for these model 
tracks and for several others are given in Table 2, where $\Phi_{\rm H}$, 
$\log T_{\rm e}$ and $\log (L/L_{\sun})$ at the blue points are also given.

\placefigure{fig6}

\placetable{table2}

A linear fit to the properties of models at the bluest points in the
HR diagram gives:
\begin{equation}
         \log \dot{M}_{\rm low} (M_{\sun} {\rm yr}^{-1}) = 
                   2.073 {M_{\rm WD}\over M_{\sun}}-8.639,
\end{equation}
as the lower boundary of the region in which steady state accretion 
solutions exist over the mass range $0.52 \le M_{\rm WD}/M_{\sun} \le 0.68$.

An approximation to the upper boundary of the region where steady state 
solutions exist can also be derived from the information in Figure 6. The 
``kink'' in each curve near the red end of each curve for the three largest 
white dwarf masses in Figure 6 actually defines the point beyond which 
static solutions are of the red giant variety, with the accretion rate 
being larger than the rate at which nuclear burning consumes fuel (Fujimoto 
1982 a,b). In Table 3, the values of $\dot{M}_{\rm H-sh}$ for several models 
is shown, along with the values of $\Phi_{\rm H}$, $\log T_{\rm e}$ and 
$\log (L/L_{\sun})$ at the kink. A linear fit between the maximum allowed 
accretion rate and the WD mass at the kinks provides
\begin{equation}
      \log \dot{M}_{\rm high} (M_{\sun} {\rm yr}^{-1})
                = 1.512\ {M_{\rm WD}\over M_{\sun}}-7.800.
\end{equation}
This line defines the upper boundary  of the region in which steady state 
accretion can occur over the white dwarf mass range $0.52 \le M_{\rm WD}
/M_{\sun} \le 0.68$. 

\placetable{table3}

The method adopted here to estimate the boundaries of the steady burning 
zone was introduced by Fujimoto (1982 a,b), who studied the properties of 
the hydrogen-burning shell in accreting models using an analytical 
solution for the envelope. He found (see Fig. 4 in Fujimoto 1982b) 
the border lines to be parallel for white dwarf masses over the range 
0.5-1.5 \msunend, while our border lines have different slopes. It is
probable that our results differ because of the approximations
used in the analytical solution. The fact that our lower border in the
$M_{\rm WD}$-$\dot{M}_{\rm H}$ plane is steeper than the upper border
is consistent with other estimates in the literature (see, e.g. Fig. 7
in Iben \& Tutukov 1996 and Fig. 10 in CIT).


\section{The Thermal Behavior of the Hydrogen-Burning Shell as a 
Function of Metallicity}

\noindent
To investigate the dependence on metallicity of the behavior of 
accreting white dwarfs, we have computed two additional sets of 
models in which hydrogen-rich matter characterized, respectively, by 
$Z=0.001$ and $Z=0.0001$ is accreted onto the same initial model of 
mass $0.516 M_{\sun}$ at the same rate $\dot{M}=10^{-8}$ \msun 
\yrm1. To make more meaningful the comparison between models accreting 
mass with different metallicities, we have adopted for all sets of
models the same helium abundance: $Y=0.28$.

In the HR diagram of Figure 7 are shown the paths during one pulse 
cycle of models of the three different metallicities. In all three 
cases, the total mass of the model is $M \sim 0.5236 M_{\odot}$. 
Several characteristics of the models are given 
in Table 4. The dependence on the metallicity of various path 
characteristics can be understood relatively 
simply. At the very lowest luminosities, all paths converge because 
the models adopt the essentially metal-independent radius of a cold 
white dwarf. Because of smaller CNO abundances, the temperatures and 
densities at the base of the accreted layer (see Table 4) must be 
larger in models of lower metallicity in order for a CNO cycle 
thermonuclear runaway to be initiated: in order to achieve larger 
densities and temperatures, more mass must be accumulated by the
lower metallicity models. This is why the time between pulses is larger,
the lower the metallicity. 
During the transition between the low and high states, the radius of 
the expanding envelope is larger, the larger the mass of the 
envelope, and this explains why, at any given luminosity, the lower 
the metallicity, the redder the model. The fact that the reddest 
point along a path is bluer, the lower the metallicity, can be accounted 
for as an envelope-opacity effect. Finally, during the plateau phase 
and during the cooling phase, the fact that, at any luminosity, the 
model of lower metallicity is redder, is again ascribable to the 
larger mass of the hydrogen-rich envelope and the consequent larger 
radius of the envelope. 

\placetable{table4}

\placefigure{fig7}

\placefigure{fig8}

In Figure 8 we have reported for comparison the evolution of 
the main physical quantities for the hydrogen-burning shell for the cases 
with $Z=0.001$ and $Z=0.0001$. The evolution of $\Phi_{\rm H}$ in panel (d) of 
this figure demonstrates graphically how the durations of both the high 
state (plateau phase) and the low state (cooling phase) increase with 
decreasing metallicity. As we have argued, the plateau phase lasts 
longer, the lower the metallicity, because the duration of the 
cooling phase and therefore the mass of hydrogen-rich material 
accreted between thermonuclear runaways increases with decreasing 
metallicity. That the amount of mass accreted during the low phase 
increases with decreasing metallicity is also evident by analyzing 
panel (a) which show that, the smaller the metallicity, the greater 
is the amount of fuel burned during the plateau phase. Figure 9
emphasizes this point once again.

\placefigure{fig9}

It is evident from Figure 7, and the discussion in \S 4, that the 
band in the $M_{\rm WD} - \dot{M}$ plane where steady state burning 
solutions exist drops to lower $\dot{M}$ as metallicity decreases. 
The semianalytical relations obtained in \S 4 suggest that a $\sim 
0.52$ \msun white dwarf accreting hydrogen-rich matter with 
$Z=0.02$ at $8\times 10^{-8}$ \msun \yrm1 settles into a steady 
state configuration, while, for an accretion rate of $2\times 
10^{-8}$ \msun \yrm1, it experiences recurrent mild flashes. To 
explore quickly the effect of the choice of metallicity on the 
location of the steady state band, we have calculated models of 
initial mass $M_{\rm WD}=0.516 M_{\sun}$ and accretion rates 
$2\times 10^{-8}$ and $8\times 10^{-8}$ \msun \yrm1, for 
metallicities of $Z=0.001$ and $Z=0.0001$. The model accreting 
hydrogen at $8\times 10^{-8}$ \msun \yrm1 settles into a red giant 
configuration after only one pulse, while the model accreting at 
$\dot{M}=2\times 10^{-8}$ \msun \yrm1 settles into a steady state 
accretion configuration after one pulse.

Adopting the method outlined in \S 4, we have estimated the limits of the 
steady burning band for the three metallicities when $M_{\rm 
WD}\simeq 0.5236$ \msunend, obtaining the values listed in Table 5.

\placetable{table5}

We have followed the long term evolution of the low $Z$ models, and from
the results (which will be described in detail elsewhere), we have
estimated the upper and lower bounds of the steady state band (see
Tables 6 and 7, respectively). 

\placetable{table6}

\placetable{table7}

The upper boundary may be approximated by
\begin{equation}
\log(\dot{M}_{\rm high}(M_{\sun} {\rm yr}^{-1}))=
2.235 \ M_{\rm WD}/M_{\sun}-8.350\ \ \ Z=0.0001
\end{equation}
\begin{equation}
\log(\dot{M}_{\rm high}(M_{\sun} {\rm yr}^{-1}))=
1.832 \ M_{\rm WD}/M_{\sun}-8.043\ \ \ Z=0.001
\end{equation} 
and the lower boundary can be approximated by

\begin{equation}
\log(\dot{M}_{\rm low}(M_{\sun} {\rm yr}^{-1}))=
3.847 \ M_{\rm WD}/M_{\sun}-9.874\ \ \ \ \ \ \ \ Z=0.0001
\end{equation} 
\begin{equation}
\log(\dot{M}_{\rm low}(M_{\sun} {\rm yr}^{-1}))=
2.969 \ M_{\rm WD}/M_{\sun}-9.236\ \ \ \ \ \ \ \ Z=0.001
\end{equation}


\section{Summary and Conclusions}

We have investigated and discussed in detail the evolutionary behaviour
of a white dwarf accreting hydrogen-rich matter of three different
metallicities: $Z=0.02, 0.001, \hbox{and } 0.0001$.

An analysis of the evolutionary behavior of several physical
characteristics of the models has shown that, for fixed values of
$M_{WD}$ and $\dot{M}$, lowering the metallicity causes the recurrence
period to become longer because, in order to achieve the larger
temperatures and densities necessary to offset the reduction of CNO
catalysts in the accreted matter, the thickness of the hydrogen-rich
accreted layer must increase.

For the steady-state burning regime, we have been able to derive borders
in the $M_{WD}$-$\dot{M}$ plane as they depend on the metal content of
the accreted matter. In agreement with earlier estimates, we find that
the area of the region in this plane in which steady-state burning takes
place becomes narrower as the white dwarf mass is increased. In addition,
the location and the extension of the steady-state burning regime have
been found to depend critically on the metallicity of the accreted matter,
as shown clearly in Figure 10, where the topology of the steady-state
region in the $M_{WD}-\dot{M}$ plane is provided for the three
metallicities considered. Reducing the metallicity, the steady-state
burning region drops to smaller accretion rates and its extension is
drastically decreased. 

The consequences of our results for the final behavior of real low
metallicity accretors are not easy to predict. On the one hand, as
metallicity is decreased, the hydrogen-burning shell becomes hotter.
This means that the underlying helium-burning layer is hotter and
less degenerate when a helium-burning thermonuclear runaway is initiated.
On the other hand, the fact that, for fixed core mass and accretion rate, 
the power of hydrogen-burning flash decreases as the metallicity is
reduced suggests that low metallicity accretors may experience relatively
mild helium shell flashes for a range of helium layer masses more extended
than in the solar metallicity case.

However, as extensively discussed in Piersanti et al.(1999), in the mild 
pulse regime, there is a parameter region in which the effects of the
hydrogen-burning shell on the helium layer are negligible, and a model
with characteristics in theis region behaves as if pure helium is
accreted. These models lead to a sub-Chandrasekhar explosion if the
initial mass of the white dwarf and the accretion rate are within a 
given range (see Fig. 1 in Tornamb\'e et al. 1998 for the solar
metallicity case). Therefore, over the long term evolution, once the 
helium-burning layer becomes thermally decoupled from the hydrogen-burning
shell, the accreted layer behaves in a way that depends only on
the accretion rate and not on the metallicity. 

Due to the prohibitively long computing time required, we have only
partially studied the long term evolution of models accreting
hydrogen-rich matter of metallicities $Z=0.001$ and $Z=0.0001$ at the
mass-accretion rate of $\dot{M}=10^{-8}$ \msun\yrm1. Both models show
that the helium layer and the hydrogen-rich layer become decoupled
as in the case  of accretion of hydrogen-rich matter of solar metallicity
(Piersanti et al. 1999). Such models will likely experience similar
outcomes independent of the metallicity of the hydrogen-rich accreted matter. 

On the basis of the results obtained so far, we suggest that, on lowering
the metallicity, the area in the $M_{WD} - \dot{M}$ plane suitable for
sub-Chandrasekhar dynamical outcomes is shifted toward slightly lower
values of $\dot{M}$, remaining  almost unchanged in extention, as indicated 
in Tornamb\'e et al. (1998) for the solar metallicity.
It has to be finally considered that metallicity could even  
affect other parameters of the binary system (as, for instance, initial 
white dwarf masses, accretion rates, etc) with the consequence that in
the real world, this scenario could be also significantly changed.


\newpage
\figcaption[] {The track in the HR diagram of a white dwarf of 
mass $\sim 0.5236$\msun accreting hydrogen-rich matter of composition 
$Y = 0.28$, $Z = 0.02$ at the rate $\dot{M} = 10^{-8}$ \msun \yrm1. 
Evolution progresses in a counter clockwise fashion along the track from the 
reddest point (RP) to the bluest point (BP) and so on. The labels IC and EC 
indicate, respectively, the onset and offset of shell convection driven by 
a hydrogen shell flash. 
$\Phi_{\rm H} = L_{\rm H}/L_{\rm s}$ and $\Phi_{\rm He} = L_{\rm He}/L_{\rm s}$
where $L_{\rm H}$, $L_{\rm He}$, and $L_{\rm s}$ are respectively, the 
hydrogen-burning, helium-burning, and surface luminosities.
In the plot we have also indicated selected specific models 
for which thermal and nuclear burning characteristics are displayed in 
Figs. 3 and 4 (see text).\label{fig1}}

\figcaption[] {The evolution during two successive pulse cycles of several 
characteristics of the model shown in Fig. 1. In panel (a), the dashed 
line gives the total mass of the model and the solid line gives the mass in 
the hydrogen-burning shell where the rate of nuclear energy generation 
$\epsilon_{\rm H}$ is at a maximum. Panels (b) and (c) give, respectively, 
the temperature and density at the point where $\epsilon_{\rm H}$ is at a 
maximum. In panel (d), $\Phi_{\rm gr} = L_{\rm gr}/ L_{\rm s}$, $L_{\rm gr}$ 
is the rate of release of gravothermal energy, and $L_{\rm s}$ and 
$\Phi_{\rm H}$ are defined in the caption of Fig. 1. Panel (e) gives the 
radius of the point where $\epsilon_{\rm H}$ is at a maximum and panel (f) 
gives the radius of the surface. \label{fig2}}

\figcaption[] {The evolution of temperature profiles over the outer part of 
the accreting white dwarf which follows the track in the HR diagram given in 
Fig. 1. Every curve corresponds to one of the models indicated in Fig. 1 
(see text).\label{fig3}}


\figcaption[] {The evolution of the profiles of the H-burning efficiency 
($\epsilon_{\rm n}$) 
over the outer part of the accreting white dwarf which follows the track in 
the HR diagram given in Fig. 1. Every curve corresponds to one of the models 
indicated in Fig. 1, as described in the text.\label{fig4}}


\figcaption[] {The rate of energy generation at its maximum in the 
hydrogen-burning shell (panel a) and the rate at which the center of the 
hydrogen-burning shell processes mass (panel b) along the high luminosity 
branch for the model with $M\sim 0.5236$ \msun and $\dot{M}=10^{-8}$ \msun 
\yrm1 for the solar metallicity case.\label{fig5}}


\figcaption[] {The rate at which the hydrogen-burning shell processes matter 
along the high luminosity branch for four different white dwarf masses as
labelled. 
Hydrogen-rich matter with $Y=0.28$ and $Z=0.02$ is accreted at the rate 
$10^{-8}$ \msun \yrm1. \label{fig6}}


\figcaption[] {Evolution in the HR diagram during one hydrogen-pulse cycle 
for white dwarf models accreting hydrogen-rich matter with three 
different metallicities: $Z = 0.02, 0.001$ and $0.0001$. The three models 
have almost the same mass ($M \sim 0.523 M_{\sun}$). \label{fig7}}


\figcaption[] {The same as in Figure 2, but for the cases Z=0.001 (solid line)
and Z=0.0001 (heavy dashed line) (see text).\label{fig8}}


\figcaption[] {Evolution of the mass coordinate of the center of the 
hydrogen-burning shell (where the rate of energy generation is at a maximum) 
in models which accrete hydrogen-rich matter with metallicities of $Z = 0.02, 
0.001$, and $0.0001$ at the rate $\dot{M}=10^{-8}$ \msun \yrm1. \label{fig9}}


\figcaption[] {The steady burning zone in the $M_{\rm WD} - \dot{M}$ plane for
the three metallicity (Z=0.02, Z=0.001 and Z=0.0001), as obtained in the 
present work. \label{fig10}}


\newpage


\begin{deluxetable}{lcc}
\tablecaption{Selected evolutionary and structural properties of the initial 
model and of the structure after it has experienced about 60 H-pulses.
\label{table1}}
\tablehead{
\colhead{} & 
\colhead{\bf Initial model} & 
\colhead{\bf Model after 60 H-pulses\tablenotemark{1)}} 
}
\startdata
Age ($10^{8}\ {\rm} yr$)          & 1.4488 & 1.4740 \nl
M (\msun)                         & 0.5168 & 0.5235 \nl
$\log(L/L_\odot)$                 & -3.770 & 3.421  \nl
$\log(T_{\rm e})$                 & 3.7479 & 5.324  \nl
$\log(R/R_\odot)$                 & -1.858 & -1.416 \nl
$M_{\rm H-sh}$ (\msun)            & 0.5163 & 0.5232 \nl
$M_{\rm He-sh}$ (\msun)           & 0.4793 & 0.4793 \nl
$\log(T_{\rm H-sh})$              & 6.5666 & 7.7300 \nl
$\log(\rho_{\rm H-sh})$           & 3.8968 & 1.5815 \nl
$\log(T_{\rm He-sh})$             & 6.5823 & 7,7600 \nl
$\log(\rho_{\rm He-sh})$          & 5.2824 & 3.4481 \nl
$\log(T_{\rm c})$                 & 6.6017 & 6.9355 \nl
$\log(\rho_{\rm c})$              & 6.3700 & 6.4091 \nl
\enddata
\nl

\tablenotetext{1)}{The listed quantities refer to the model at the bluest 
point along the loop in the HR diagram.}

\end{deluxetable}

\newpage


\begin{deluxetable}{ccccc} 
\tablecaption{Minimum accretion rate for steady burning accretion as a 
function of white dwarf mass. The chemical composition of the accreted 
matter is: $Y = 0.28$, $Z = 0.02$. 
The location in the HR diagram of the bluest point along the track
($\log T_{\rm e}$, $\log L/L_{\sun}$) and 
$\Phi_{\rm H} = L_{\rm H}/L_{\sun}$ are also reported.\label{table2}} 
\tablehead{
\colhead{$M_{\rm WD}$(\msunend)} & 
\colhead{$\log(\dot{M})$(\msun \yrm1)} & 
\colhead{$\Phi_{\rm H} $} & 
\colhead{$\log T_{\rm e}$} & 
\colhead{$\log(L/L_{\sun}$)} 
}
\startdata
0.520 & -7.585 & 0.939 & 5.325 & 3.421 \nl
0.540 & -7.520 & 0.952 & 5.334 & 3.562 \nl
0.560 & -7.493 & 0.938 & 5.358 & 3.519 \nl
0.581 & -7.428 & 0.942 & 5.380 & 3.583 \nl
0.600 & -7.378 & 0.938 & 5.401 & 3.634 \nl
0.620 & -7.345 & 0.930 & 5.423 & 3.671 \nl
0.640 & -7.336 & 0.914 & 5.444 & 3.688 \nl
0.660 & -7.255 & 0.942 & 5.460 & 3.755 \nl
0.680 & -7.236 & 0.935 & 5.479 & 3.777 \nl
\enddata
\nl
\end{deluxetable}

\newpage


\begin{deluxetable}{ccccc}
\tablecaption{Maximum accretion rate for steady state burning and position 
in HR diagram as a function of white dwarf mass. Composition of accreted 
matter is the same as in Table 1. The discontinuity in 
$\log T_{\rm e}$ at $M_{\rm WD} > 0.6$ \msun is due to a different approach 
in the treatment of atmospheric layers adopted to prevent a huge expansion 
of the envelope.\label{table3}}
\tablehead{
\colhead{$M_{\rm WD}$ (\msunend)} & 
\colhead{$\log(\dot{M})$ (\msun \yrm1)} & 
\colhead{$\Phi_{\rm H}$} & 
\colhead{$\log T_{\rm e}$} & 
\colhead{$\log(L/L_{\sun})$} 
}
\startdata
0.520 & -7.018 & 1.042 & 4.474 & 3.949 \nl
0.540 & -6.988 & 1.044 & 4.544 & 3.978 \nl
0.560 & -6.958 & 1.024 & 4.434 & 4.014 \nl
0.580 & -6.917 & 1.031 & 4.340 & 4.055 \nl
0.600 & -6.894 & 1.005 & 4.236 & 4.088 \nl
0.620 & -6.859 & 1.004 & 4.642 & 4.124 \nl
0.640 & -6.821 & 1.005 & 4.583 & 4.161 \nl
0.660 & -6.807 & 1.006 & 4.896 & 4.174 \nl
0.680 & -6.779 & 1.006 & 4.923 & 4.203 \nl
\enddata
\nl
\end{deluxetable}

\newpage


\begin{deluxetable}{lccc}
\tablecaption{The principal characteristics of the helium- and 
hydrogen-burning shells as a function of metallicity $Z$ for models of mass 
$M\sim 0.5236$ \msun accreting hydrogen-rich matter at the rate $\dot{M} = 
10^{-8}$ \msun \yrm1.\label{table4}}
\tablehead{
\colhead{} &
\colhead{$Z=0.02$} & 
\colhead{$Z=0.001$} & 
\colhead{$Z=0.0001$}
}
\startdata
{Period}\tablenotemark{a)} & 1.24 & 2.39 & 3.27 \nl 
{$L_{\rm H,max}$}\tablenotemark{b)} & 13.80 & 8.62 & 2.61 \nl
{$L_{\rm H,min}$}\tablenotemark{c)} & 0.394 & 0.650 & 1.450 \nl
{$L_{\rm He,max}$}\tablenotemark{d)} & 1.75 & 4.70 & 56.20 \nl
{$\Delta M_{\rm H, max}$}\tablenotemark{e)} & 1.76 & 3.45 & 5.04 \nl
{$\Delta M_{\rm H, min}$}\tablenotemark{f)} & 5.70 & 12.70 & 23.20 \nl
{$\log(T_{\rm H,min})$}\tablenotemark{g)} & 7.385 & 7.436 & 7.496 \nl
{$\log(T_{\rm H,max})$}\tablenotemark{h)} & 7.953 & 7.990 & 8.005 \nl
{$\log(\rho_{\rm H,min})$}\tablenotemark{i)} & 1.214 & 1.337 & 1.559 \nl
{$\log(\rho_{\rm H,max})$}\tablenotemark{j)} & 2.734 & 2.956 & 3.067 \nl
\enddata
\nl

\tablenotetext{a)} {Time between two successive H-flash ($10^4$yr).}
\tablenotetext{b)} {Maximum luminosity of H-shell ($10^{5}L_{\sun}$).}
\tablenotetext{c)} {Minimum luminosity of H-shell ($L_{\sun}$).}
\tablenotetext{d)} {Maximum luminosity of He-shell ($10^{-6} L_{\sun}$).}
\tablenotetext{e)} {Thickness in mass of H-shell at $L_{\rm H,max}$
($10^{-4} M_{\sun}$).}
\tablenotetext{f)} {Thickness in mass of H-shell at $L_{\rm H,min}$ 
($10^{-5} M_{\sun}$).}
\tablenotetext{g)} {Temperature of H-shell at $L_{\rm H,min}$ }
\tablenotetext{h)} {Temperature of H-shell at $L_{\rm H,max}$ }
\tablenotetext{i)} {Density of H-shell at $L_{\rm H,min}$ }
\tablenotetext{j)} {Density of H-shell at $L_{\rm H,max}$ }
\end{deluxetable}

\newpage


\begin{deluxetable}{ccc}
\tablecaption{The minimum and maximum values of the accretion rate for which 
a white dwarf of mass $M \sim 0.5236$ \msun burns hydrogen as rapidly as it 
accretes it at different metallicities. \label{table5}} 
\tablehead{
\colhead{$Z$} & 
\colhead{$\dot{M}_{\rm low}$ ($10^{-8}$ \msun \yrm1)} &
\colhead{$\dot{M}_{\rm high}$ ($10^{-8}$ \msun \yrm1)} 
}
\startdata
0.0001 & 1.35 & 6.50 \nl
0.001  & 2.05 & 8.30 \nl
0.02   & 2.60 & 9.60 \nl
\enddata
\nl
\end{deluxetable}

\newpage


\begin{deluxetable}{ccccc} 
\tablecaption{Maximum accretion rate for steady state burning and position in 
the HR diagram as a function of white dwarf mass. The chemical composition 
of accreted matter is $Y = 0.28$ and $Z=0.0001$ or $Z=0.001$. \label{table6}}

\tablehead{
\colhead{$M_{\rm WD}$(\msunend)} & 
\colhead{$\log(\dot{M}$)(\msun \yrm1)} & 
\colhead{$\Phi_{\rm H} $} & 
\colhead{$\log T_{\rm e}$} & 
\colhead{$\log(L/L_{\sun})$}
}
\startdata
      &        & $Z=0.0001$ &    &     \nl
0.520 & -7.187 & 1.083 & 4.835 & 3.762 \nl
0.540 & -7.142 & 1.125 & 4.890 & 3.791 \nl
0.560 & -7.100 & 1.122 & 4.899 & 3.834 \nl
0.580 & -7.058 & 1.062 & 4.863 & 3.900 \nl
0.591 & -7.025 & 1.101 & 4.886 & 3.918 \nl
0.600 & -7.005 & 1.104 & 4.892 & 3.937 \nl
      &        & $Z=0.001$ &    &      \nl
0.520 & -7.084 & 1.099 & 4.706 & 3.859 \nl
0.540 & -7.060 & 1.073 & 4.704 & 3.894 \nl
0.560 & -7.022 & 1.086 & 4.751 & 3.927 \nl
0.580 & -6.978 & 1.084 & 4.731 & 3.971 \nl
0.590 & -6.960 & 1.089 & 4.707 & 3.987 \nl
\enddata
\nl
\end{deluxetable}

\newpage


\begin{deluxetable}{ccccc} 
\tablecaption{Minimum accretion rate for steady state burning and position in 
the HR diagram as a function of white dwarf mass for two different 
assumptions on the metallicities of the accreted matter and the same He 
content.\label{table7}}

\tablehead{
\colhead{$M_{WD}(M_{\sun})$} & 
\colhead{$\log(\dot{M})$(\msun \yrm1)} & 
\colhead{$\Phi_{\rm H}$} & 
\colhead{$\log T_{\rm e}$} & 
\colhead{$\log(L/L_{\sun}$)}
}
\startdata
      &        & $Z=0.0001$ &    &     \nl
0.520 & -7.870 & 0.967 & 5.206 & 3.142 \nl
0.540 & -7.795 & 0.966 & 5.235 & 3.204 \nl
0.560 & -7.721 & 0.964 & 5.259 & 3.279 \nl
0.580 & -7.639 & 0.963 & 5.284 & 3.362 \nl
0.591 & -7.598 & 0.963 & 5.295 & 3.402 \nl
0.600 & -7.519 & 0.968 & 5.304 & 3.479 \nl
      &        & $Z=0.001$ &    &      \nl
0.520 & -7.691 & 0.964 & 5.268 & 3.310 \nl
0.540 & -7.631 & 0.965 & 5.279 & 3.369 \nl
0.560 & -7.583 & 0.961 & 5.302 & 3.418 \nl
0.580 & -7.510 & 0.961 & 5.325 & 3.491 \nl
0.590 & -7.482 & 0.960 & 5.338 & 3.520 \nl
\enddata
\nl
\end{deluxetable}

\end{document}